\documentclass[useAMS,usenatbib]{mn2e}

\usepackage{graphicx}

\title[Bias in the Estimation of Global Luminosity Functions]{Bias in the Estimation of Global Luminosity Functions} \author[O.~Ilbert et
al.]{O.~Ilbert$^{1}$\thanks{E-mail: Olivier.Ilbert@oamp.fr},
  L.~Tresse$^{1}$, S.~Arnouts$^{1}$, E.~Zucca$^{2}$,
  S.~Bardelli$^{2}$, G.~Zamorani$^{2}$, \newauthor C.~Adami$^{1}$,
  A.~Cappi$^{2}$, B.~Garilli$^{4}$, O.~Le~F\`evre$^{1}$,
  D.~Maccagni$^{4}$, B.~Meneux$^{1}$, \newauthor R.~Scaramella$^{5}$,
  M.~Scodeggio$^{4}$,
  G.~Vettolani$^{3}$, A.~Zanichelli$^{3}$ \\
  $^{1}$Laboratoire d'Astrophysique de Marseille, Les Trois-Lucs, B.P.~8,
  13376, Marseille Cedex 12, France \\
  $^{2}$INAF, Osservatorio Astronomico di Bologna, via Ranzani 1, 40127 Bologna, Italy\\
  $^{3}$Istituto di Radioastronomia-CNR, via Gobetti 101, 40129 Bologna, Italy \\
  $^{4}$Istituto di Astrofisica Spaziale e Fisica Cosmica del CNR, via Bassini 15, 20133 Milano, Italy\\
  $^{5}$Ossevatorio Astronomico di Roma, Via Ossevatorio 2, 00040
  Monteporzio Carone (Roma), Italy}

\begin{document}

\date{Accepted. Received; in original form}

\pagerange{\pageref{firstpage}--\pageref{lastpage}} \pubyear{2004}

\maketitle
 
\label{firstpage}

\begin{abstract}   
  
  We discuss a bias present in the calculation of the global
  luminosity function (LF) which occurs when analysing faint galaxy
  samples.  This effect exists because of the different spectral
  energy distributions of galaxies, which are in turn quantified by
  the $k$-corrections. We demonstrate that this bias occurs because
  not all galaxy types are visible in the same absolute magnitude
  range at a given redshift and it mainly arises at high redshift
  since it is related to large $k$-corrections. We use realistic
  simulations with observed LFs to investigate the amplitude of the
  bias.  We also compare our results to the global LFs derived from
  Hubble Deep Field-North and -South (HDF) surveys.  We conclude that,
  as expected, there is no bias in the global LF measured in the
  absolute magnitude range where all galaxy types are observable.
  Beyond this range the faint-end slope of the global LF can be
  over/under-estimated depending on the adopted LF estimator. The
  effect is larger when the reference filter in which the global LF is
  measured, is far from the rest-frame filter in which galaxies are
  selected.  The fact that LF estimators are differently affected by
  this bias implies that the bias is minimal when the different LF
  estimators give measurements consistent with one another at the
  faint-end.  For instance, we show that the estimators are discrepant
  in the same way both in the simulated and HDF LFs. This suggests
  that the HDF LFs are affected by the presently studied bias.  The
  best solution to avoid this bias is to derive the global LF in the
  reference filter closest to the rest-frame selection filter.

\end{abstract}


\section{INTRODUCTION}

The luminosity function (LF) is a fundamental and basic tool to understand and
constrain the history of galaxy formation and evolution.  Moreover, the
derived mean luminosity density at different redshifts allows to derive
estimates of the cosmic star formation density.  In the distant Universe, LFs
are measured in se\-ve\-ral redshift bins in order to quantify
the evolution of galaxy populations.  In this paper, we focus on the
relia\-bi\-lity of the statistical estimators usually used to measure the
global LF.  We call {\it global LF} the sum of the LFs per galaxy type
\citep{1988ARA&A..26..509B}.  Calculating LFs is not a trivial task since
estimators must account for all biases or limits introduced by the
observational selection effects. Most of the surveys are limited in apparent
magnitude. This effect is accounted for in the 1/V$_{\rm max}$ LF estimator
\citep{1968ApJ...151..393S}.  The drawback of the 1/V$_{\rm max}$ method is
the implicit assumption, in its formulation, of a uniform galaxy distribution
(i.e. no significant over- or under-densities of galaxies).
Nevertheless, because of its simplicity, this method is the most often used in
high-redshift surveys.  \citet{1971MNRAS.155...95L} developed the C$^{-}$
method to overcome the assumption of a uniform galaxy distribution.  The STY
\citep*{1979ApJ...232..352S} and the Step Wise Maximum-Likelihood LF
estimators, hereafter SWML, \citep*[][EEP]{1988MNRAS.232..431E} are both
related to maximum-likelihood statistical methods.  
The C$^{-}$, STY and SWML methods make no assumptions on
  spatial distribution of galaxies, but the information about the
  normalization of the LF is lost. \citet{1982ApJ...254..437D}
  reviewed various estimators to derive the normalization.
  In contrast to C$^{-}$ and SWML, the STY method does
  assume a parametric form to the luminosity distribution.

All LF estimators present both advantages and drawbacks.
\citet{1997AJ....114..898W} and \citet*{2000ApJS..129....1T} compared
several LF estimators using simulated ca\-talogues.  Their mock
catalogues did not tackle into detail the effects of $k$-corrections
and of the mix of individual and different LF shapes for different
morphological types in the measurement of the global LF. The
Canada-France Redshift Survey \citep[CFRS;][]{1995ApJ...455..108L}
demons\-trated that the evolution of the LF depends strongly on the
studied galaxy population. In this paper we add the dependency of limiting
absolute magnitudes on galaxy type in si\-mu\-lated catalogues, and at
the same time we introduce an evolution of the LF per galaxy
population to produce realistic simulations. 
These improvements enable us to identify an intrinsic bias in the
  estimators to measure the global LF.
  
  The different visibility limits for the various galaxy types (mainly
  due to different $k$-corrections) affects all flux-limited surveys.
  Hence it can have an impact on statistical analyses, in particular
  the LF estimates.  The accepted idea is that certain galaxy types
  sometimes can not be visible in a given redshift bin, so that it
  would ob\-vi\-ous\-ly underestimate the global LF.  However even
  though all galaxy types are visible in a given redshift bin, we show
  using realistic simulations that a bias still arises in the
  measurement of the global LF.  As noted by
  \citet{1995ApJ...455..108L}, it occurs because different galaxy
  types are not visible in the same absolute magnitude range.  In the
  literature this bias has never been quantified. We use real and
  simulated data to investigate the amplitude and the behavior of this
  bias. In particular our analysis is focused on high-redshift data
  since the $k$-correction values are small at low redshift.

One possible solution to avoid this bias would be to sum the
extrapolated LFs per galaxy type to measure the global LF.
Unfortunately this solution is hazardous in the highest redshift bins
of a deep survey for two reasons: the number of galaxies is often too
small to derive LFs per galaxy type and not all the LF slopes per
galaxy type are well constrained, which would imply a dangerous
extrapolation.  The analysis of the global high-redshift LFs has been
the framework of most of the previous analyses on deep surveys like,
for instance, the Subaru Deep Field
\citep[SDF;][]{2003AJ....125...53K}, the Hubble Deep Fields (HDF; e.g.
for example \citealt*{1997AJ....113....1S},
\citealt{2000ApJS..129....1T}, \citealt*{2002A&A...395..443B}), the
CFRS \citep{1995ApJ...455..108L}.  With the on-going or earlier deep
surveys, one needs to quantify in details this bias related to
$k$-correction effects.

This paper is organized as follows.  Section~2 describes the origin of
the bias linked to the spectral energy distribution dependency of
absolute magnitudes. Section~3 reviews briefly the following
estimators, 1/V$_{\rm max}$, STY, SWML and C$^{+}$, and the bias
linked to each of them. Section~4 quantifies the impact of the bias on
the global LF from simulations and from the Hubble Deep Field surveys.
Section~5 presents our conclusion.  Throughout this paper, we adopt an
Einstein-de Sitter universe ($\Omega_{\rm 0} = 1$, $\Omega_{\rm
  \Lambda} = 0$) and H$_{\rm 0} = 100$~km~s$^{-1}$~Mpc$^{-1}$, but the
results here discussed are not dependent on the adopted cosmological
model.

\section{ORIGIN OF THE BIAS} 

In any flux-limited survey, galaxy $i$ is observed within a fixed
apparent magnitude range, $m^{S}_{bright} \leq m^{S}_i \leq
m^{S}_{faint}$, where $S$ designates the band where galaxies are selected.
We use $Ref$ to designate the reference
filter corresponding to the wavelength at which the LF is estimated.
The observational limits imply that galaxy $i$ with spectral energy
distribution SED$_i$ is observable within a fixed redshift range and a
fixed absolute magnitude range.  

\begin{figure}
\includegraphics[width=8.5cm]{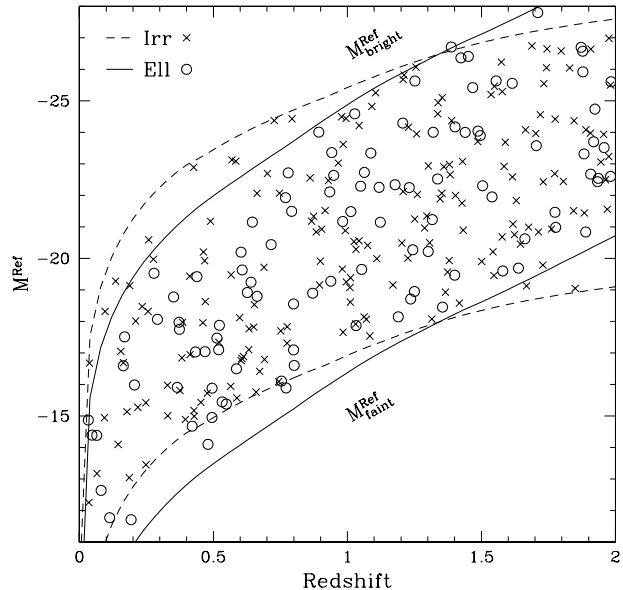} 
\caption{Example of the observable $M-z$ plane in a flux-limited survey for
  elliptical-type SEDs (open circles) limited by the solid lines, and
  irregular-type SEDs (crosses) limited by the dashed lines.  Each type of
  galaxies is observable between its respective limits of absolute magnitudes,
  $M^{Ref}_{bright}$ and $M^{Ref}_{faint}$.  The $M-z$ planes for different
  SEDs do not coincide. In this example the selection filter is the I band
  and the absolute magnitudes are computed in the U band.}
\label{fig1}
\end{figure}

Obviously, the redshift and
absolute magnitude limits depend on SED$_i$ (see Fig.~\ref{fig1}).
That is, galaxy SED$_i$ with absolute magnitude, $M_i^{Ref}$, is
observable in the redshift range:
$$
z_{bright}(M_i^{Ref}, {\rm SED}_i) \leq z_i \leq z_{faint}(M_i^{Ref}, {\rm SED}_i).
$$
Similary galaxy SED$_i$ at redshift $z_i$ is observable in the
 absolute magnitude range:
$$
M_{bright}^{Ref}(z_i, {\rm SED}_i) \leq M_i^{Ref} \leq M_{faint}^{Ref}(z_i, {\rm SED}_i).
$$
The latter limits of absolute magnitudes are defined by,
\begin{eqnarray}\label{eqn:Mf}
M^{Ref}_{bright}(z, {\rm SED}_i)& = & m^{S}_{bright} - DM(z) - KC(z,{\rm SED}_i), \nonumber \\
M^{Ref}_{faint}(z, {\rm SED}_i) & = & m^{S}_{faint}  - DM(z) - KC(z,{\rm SED}_i), \nonumber 
\end{eqnarray}
where $DM(z)$ is the distance modulus and $KC$ is expressed as follows:
$$
KC(z, {\rm SED}_i) = (k^{Ref}(z) + m^{S}(z) - m^{Ref}(z))^{{\rm SED}_i},
$$
where $k$ is the $k$-correction, and $m$ is the apparent magnitude
measured from the SED.  We note that, for typical galaxy
k-corrections, $M^{Ref}_{bright}$ and $M^{Ref}_{faint}$ are overall
strictly decreasing (i.e. increasing in brightness) as a function of
redshift.  Thus, in a given redshift interval, $z_{low} \leq z <
z_{high}$, it is impossible to observe galaxy SED$_i$ fainter than
$M^{Ref}_{faint}(z_{low}, {\rm SED}_i)$ and brighter than
$M^{Ref}_{bright}(z_{high}, {\rm SED}_i)$. Then, in a given absolute
magnitude interval, some SEDs may not be observable while others are
detected.  This leads to a bias intrinsic to the global LF estimated
in a given redshift interval, $z_{low} \leq z < z_{high}$, as we
describe in details in Section~3. 
We discuss in this paper the  
bias in the luminosity function induced by different SEDs. However 
a similar bias arises when using a magnitude limited sample 
to derive estimates of other distribution functions, like masses
or sizes.

\begin{figure}
\includegraphics[width=8.5cm]{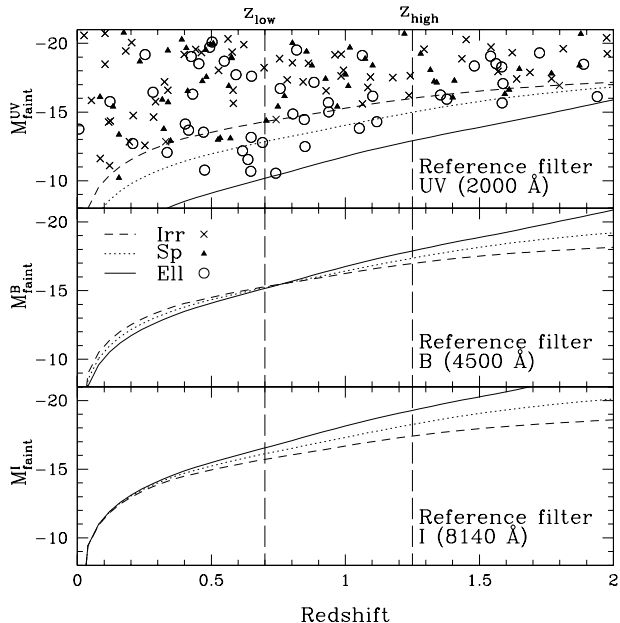} 
\caption{Faint observable absolute magnitude limits in different
  reference filters as a function of redshift for various SEDs (see
  Section~2). The filter to select galaxies is $I$ with $I_{AB} \leq
  26$ mag. The reference filters are respectively the UV FOCA filter
  (2000~\AA), the B HST filter (4500~\AA), the I HST filter
  (8140~\AA). In each panel, we show three templates from a set of
  SEDs described in Section~4.1; solid, dotted and short-dashed lines
  represent elliptical-, spiral- and irregular-type SED galaxies
  respectively.  Vertical long dashed lines define the studied redshift
  interval, $0.7 \leq z < 1.25$. In the top panel, we illustrate 
  positions of observable galaxies with open circles for 
  elliptical-type SEDs, filled triangles for spiral-type SEDs, and
  crosses for irregular-type SEDs.}
\label{fig2}
\end{figure}

Of course, no bias would be present if the entire population under
study had the same or very similar SEDs, as in the case of the
estimate of the LF for a single galaxy type.  When this is not the
case, as for example in the case of galaxies of all types (from very
blue to very red) an obvious bias is arising. As shown in
Fig.~\ref{fig2}, there are three possible cases, which are defined on
the basis of $z$, $\lambda^S$ (effective wavelength of the selection
filter) and $\lambda^{Ref}$ (effective wavelength of the reference
filter):
\begin{itemize}
\item $1+z_{low} < \lambda^S/\lambda^{Ref}$, the faint limiting absolute
  magnitude is brighter for blue galaxies than for red galaxies and
  therefore faint blue galaxies are missing from the sample;
\item $1+z_{low} \sim \lambda^S/\lambda^{Ref}$, the faint limiting
  absolute magnitude is about the same for all galaxies;
\item $1+z_{low} > \lambda^S/\lambda^{Ref}$, the faint limiting absolute
  magnitude is brighter for red galaxies than for blue galaxies and
  therefore faint red galaxies are missing from the sample.
\end{itemize}
The impact of the bias on each LF estimator depends 
on the case in which the analysis of the global LF is performed and 
on the different slopes and normalizations of red and blue galaxies.
One possible way to cope with this effect is to derive the global LF
in an absolute magnitude range in which all types are observable
as done in e.g.~\citet*{1997ApJ...487..512S}.
However this means to throw away from the analysis some fraction 
of the data.
Another way is to define the sample to be analysed in such a way
that $1+z_{low} \sim \lambda^S/\lambda^{Ref}$ for each redshift bin
\citep[see e.g.][2003]{2001ApJ...551L..45P}. This requires the availability
of a multi-color survey, if one wants to estimate the LF in a fixed 
$\lambda^{Ref}$ over a wide redshift range.

Fig.~\ref{fig2} shows the faint observable absolute magnitude limits,
$M^{Ref}_{faint}$, as a function of $z$ for three SEDs (E, Sp, Irr) in
the case of an $I$-selection filter, $m^{I}_{faint} = 26$~mag. In 
Fig.~\ref{fig2}, three reference filters are considered:
 UV(2000~\AA), B(4500~\AA) and I(8140~\AA).
For illustration purposes, we consider galaxy populations in the
redshift range $0.70 \leq z_i < 1.25$.  Let us consider the case of
the UV reference filter (top panel of Fig.~\ref{fig2}).
Irregular-type SEDs are visible in the whole $z$ interval
[$0.70, 1.25$] if they are brighter than $M^{UV} = -16.0$ (case~$a$).
They are still visible, but not in the whole $z$ interval, if $-16.0 <
M^{UV} < -14.2$ (case~$b$). Finally, irregular-type SEDs fainter than $M^{UV}
= -14.2$ mag are not visible at all in this $z$ interval (case~$c$).
In case~$a$, no correction for missing objects has to be applied when 
computing the LF.
In case~$b$, LF estimators enable to correct for the fact that some
galaxies cannot be visible in the whole redshift range.
In case~$c$, the global LF cannot be measured because irregular-type
SEDs are lost. The same considerations apply for spiral- and
elliptical-type SEDs, except that the observable $M^{UV}$ limits have
different values.  In our first example (UV reference filter), faint
galaxies of the irregular-type SEDs become unobservable in an absolute
magnitude range where galaxies with other SEDs can still be detected.
The resulting estimate of the global LF would start to be biased,
because of the absence from the sample of these galaxies, for absolute
magnitude bins fainter than $M^{UV} = -14.2$.  As shown in the top
panel of Fig.~\ref{fig2}, in this example the luminosity function for
elliptical type SEDs can be estimated down to $M^{UV} \sim -10.0$.  As
a consequence, the range of magnitudes over which the estimate of the
global LF would be biased is of the order of 4 magnitudes. Obviously,
this range is much smaller and close to zero when the rest frame
wavelength corresponding to the selection filter is close to the
reference filter (see, for example, the middle panel in
Fig.~\ref{fig2}, where Ref = B).  In case Ref = I, the biased range
spans about one magnitude (bottom panel of Fig.~\ref{fig2}), but in
this case are the early type SEDs which are missing from the sample in
an absolute magnitude range where irregular type SEDs are still
visible.

In summary, the bias we are discussing arises when estimating the LFs
of the entire galaxy population in deep surveys because of the
dependency of limiting absolute magnitudes on galaxy types.  In
estimating LFs, particular care has to be taken in the faintest
absolute magnitude bins, where some SEDs are no longer observable, thus
introducing a bias in the measurement of the LFs. There is no bias in
a the absolute magnitude range where all SEDs can be observed. In the
case where the redshift bin starts at $z_{low} = 0$, the bias does not
appear since all SEDs have the same faintest observable absolute
magnitude limit.

\section{BIAS INTRINSIC TO THE LUMINOSITY FUNCTION ESTIMATORS}

In this section, we des\-cribe the intrinsic bias which arises when one
applies the commonly used LF estimators, 1/V$_{\rm max}$, C$^{+}$, SWML and
STY to estimate the global LF of the population composed by objects
with different SEDs. We here describe the case of a magnitude
limited sample with both bright and faint 
apparent ma\-gnitude cuts, $m_{bright}^{S}$ and $m_{faint}^{S}$.  In a given
redshift interval, [$z_{low}$, $z_{high}$], the minimum and maximum observable
redshifts for galaxy~$i$ are $z_{min, i}$~=~max[$z_{low}$,
$z_{bright}(M^{Ref}_i, {\rm SED}_i)$] and $z_{max, i}$~=~min[$z_{high}$,
$z_{faint}(M^{Ref}_i, {\rm SED}_i)$].  We call $N_g$, the total number of
galaxies observed in the redshift interval, $z_{low} \leq z < z_{high}$. We
first analyse the 1/V$_{\rm max}$ and C$^{+}$ estimators which are biased in a
similar way with respect to the measurement of the global LF; then we analyse
the maximum-likelihood methods, SWML and STY.

\subsection{The 1/V$_{\rm max}$ and  C$^{+}$ estimators} 

\subsubsection{The 1/V$_{\rm max}$ estimator} 

The maximum observable comoving volume in which galaxy~$i$ can be
detected, is given by
\begin{equation} \label{eqn:Vmax}
V_{obs, i} = \int_\omega\int^{z_{max, i}}_{z_{min, i}}\frac{d^2V} {d\omega dz} d\omega dz, 
\end{equation}
where $\omega$ is the effective solid angle of the survey, and $V$ is the
comoving volume.  The LF measured in the reference filter $Ref$,
$\phi^{Ref}(M)$, is discretized in bins of absolute magnitudes with width $dM$
as follows:
\begin{equation} \label{eqn:PhiDisc}
\phi^{Ref}(M) = \sum^{N_{bin}}_{k=1} \phi_k^{Ref} W(M_k^{Ref} - M), 
\end{equation}
where the window function $W$ is defined as, 
\begin{equation} \label{eqn:W}
W(M_k^{Ref} - M) = 
\left\{
\begin{array}{rl}
1 & \mbox{if $-dM/2 \le M_k^{Ref}-M < dM/2 $} \\
0 & \mbox{otherwise,}
\end{array}
\right.
\end{equation}
and the discrete values of the LF, $\phi_k^{Ref}$, are derived in each absolute
magnitude bin $k$ as follows:
\begin{equation} \label{eqn:PhiVmax}
\phi_k^{Ref} dM = \frac{1}{V_{total}}\sum^{N_g}_{i=1}\frac{V_{total}}{V_{obs, i}}W(M_k^{Ref} - M_i^{Ref}), 
\end{equation}
where $V_{total}$ is the comoving volume between $z_{low}$ and $z_{high}$.
Although not necessary, the term $V_{total}$ is introduced to point out the
weight, $\frac{V_{total}}{V_{obs, i}}$, applied to each galaxy~$i$.  Galaxies
which belong to the non observable redshift range, [$z_{low}$, $z_{min,
i}$]$\cup$[$z_{max, i}$, $z_{high}$], contribute to $\phi^{Ref}_k$ through
this weight.  This weighting scheme assumes a homogenous galaxy distribution,
and it enables the estimator to recover the right number of galaxies with the
same SED$_i$ in the redshift bin.

\subsubsection{The C$^{+}$ estimator}

\citet{1971MNRAS.155...95L} derived the C$^{-}$ method.  We use a
mo\-di\-fied version of the C$^{-}$, called C$^{+}$
\citep{1997A&A...326..477Z}. The contribution of each galaxy $i$
to the LF in the reference filter, $Ref$, can be expressed by the following 
recursive expression:
\begin{equation} \label{eqn:phiC+cont}
\psi(M_i^{Ref}) = \frac{1-\sum_{j=1}^{i-1} \psi(M_j^{Ref})}{C^{+}(M_i^{Ref})}, 
\end{equation}
where galaxies are sorted from the faintest ($j=1$) to the brightest
absolute magnitude, and C$^+$ is the number of galaxies with
$M^{Ref} < M_i^{Ref}$ and $z_{low} \leq z < z_{max, i}$.
The cumulative luminosity function in Eq.~\ref{eqn:phiC+cont} is normalized
to unity at the minimum luminosity of the galaxies in the sample. The 
absolute normalization A is then derived using the method described in EEP88.

The LF in a given absolute magnitude bin $k$ is then the sum of the
contributions of all galaxies to this bin,
\begin{equation} \label{eqn:PhiC+}
\phi_k^{Ref} dM = A \sum_{i=1}^{N_g} \psi(M_i^{Ref}) W(M_k^{Ref} - M_i^{Ref}). 
\end{equation}
$\psi(M_i^{Ref})$ is the contribution of galaxy $i$ to $\phi^{Ref}_k$, and
this contribution enables to recover the right number of galaxies
with the same SED$_i$ in the redshift bin. 

\subsubsection{Intrinsic bias in the global 1/V$_{\rm max}$ and C$^{+}$ LFs}

Let us consider the $k^{th}$ bin of the LF, of width $dM$, centered on
$M_k^{Ref}$. Let us take the following example,
$M^{Ref}_{faint}(z_{low}, {\rm SED}_1) = M_k^{Ref} - dM/2$ and
$M^{Ref}_{faint}(z_{low}, {\rm SED}_2) < M_k^{Ref} + dM/2$.  In this
case, both SEDs are observed in bin ($k-1$). SED$_1$ and SED$_2$
galaxies contribute separately to $\phi_{k-1}^{Ref}$, and thus
$\phi_{k-1}^{Ref}$ is well reco\-ve\-red.  In bin $k$, SED$_1$
galaxies are no more observable and only SED$_2$ galaxies
contribute to $\phi_k^{Ref}$.  As a consequence, $\phi_k^{Ref}$ is
equal to the LF of SED$_2$ galaxies.  In Fig.~\ref{fig3} we show the
bias of the global LF estimated with the 1/V$_{\rm max}$ method (open
circles) and with the C$^{+}$ method (open squares), adopting
different input LFs for the two SEDs. In the upper-panel, we adopt the
same LF for the SED$_1$ late-type and SED$_2$ early-type galaxies.  In
the lower-panel, different slopes have been used for the input LFs
($\alpha=-1.6$ for late types and $\alpha=-0.5$ for early types).  In
both cases, beyond the absolute magnitude limit where late-type SEDs
are not observable, the 1/V$_{\rm max}$ and the C$^{+}$ methods recover the
slope of the remaining galaxy population.  An other way to explain the
bias which affects the 1/V$_{\rm max}$ and C$^{+}$ estimators is the
following.  If we derive the individual LFs of SED$_1$ and SED$_2$,
the estimators recover properly the LFs per type. The absolute
magnitude ranges in which the two LFs are estimated are not the same;
summing these LFs per type without extrapolating at fainter absolute
magnitudes the LF derived for SED$_1$ late-type galaxies is exactly
the same thing as directly deriving the global LF for the 1/V$_{\rm
  max}$ and C$^{+}$ estimators. It clearly appears that in the
absolute magnitude range [$-17.7, -13.5$] we are summing the LF of
the remaining type (SED$_2$).  In the case of a number of galaxies
large enough to derive LF per type, one possible solution to overcome
the bias would be to derive the global LF by summing the extrapolated
LFs per type.  We note that the C$^{+}$ normalization (using EEP88
method) is slightly overestimated.  We have checked that this
normalization method recovers the input normalization only if the LF
is estimated with SWML or STY methods, and not with the C$^{+}$
method.

In conclusion, the 1/V$_{\rm max}$ and C$^{+}$ methods lead always
to underestimating the LF in the faintest absolute magnitude
bins, thus biasing the global faint-end slope of the LF.

\subsection{The maximum likelihood estimators}

\subsubsection{The STY and SWML estimators}

The STY \citep{1979ApJ...232..352S} and the SWML (e.g. EEP88) estimators are
both derived from maximum-likelihood me\-thods.  The likelihood $\mathcal L$
is the probability to obtain a sample equal to the observed one
within the apparent magnitude limits of the survey. $\mathcal L$ is computed
as the product of the probabilities to
observe each galaxy at $M^{Ref}_{i}$,
\begin{equation} \label{eqn:Like} 
\mathcal L = \prod^{N_g}_{i=1} p(M_i^{Ref}) = \prod^{N_g}_{i=1} \frac{\phi^{Ref}(M_i^{Ref})}{\int^{M_{faint}^{Ref}(z_i, {\rm SED}_i)}_{M_{bright}^{Ref}(z_i, {\rm SED}_i)} \phi^{Ref}(M)dM},
\end{equation}
where $M^{Ref}_{faint}(z_i, {\rm SED}_i)$ and $M^{Ref}_{bright}(z_i,
{\rm SED}_i)$ are the faint and bright observable absolute magnitude
limits of galaxy~$i$ at $z_i$ (see Section~2). We maximize $\mathcal
L$ with respect to the LF.  For the STY estimator, a functional form
for the LF is assumed. We use the Schechter function
\citep{1976ApJ...203..297S}. For the SWML estimator, the LF is
discretized into absolute magnitude bins (see Eq.~\ref{eqn:PhiDisc}).
No assumption is made about the LF shape.  The STY parame\-tric and
SWML non-parametric estimators are complementary.  The error bars in
the SWML estimator are derived from the covariance matrix and the
normalization is the sum of the inverse of the selection function of
each galaxy as described in EEP88.

\subsubsection{Intrinsic bias in the global SWML and STY LFs}

The SWML and STY estimators are biased in a different way with respect to the
1/V$_{\rm max}$ and C$^+$ estimators. Let us write $\mathcal L$ as the product
of $\mathcal L_1$ and $\mathcal L_2$, which are respectively the likelihood
for SED$_1$ and SED$_2$ galaxy populations within a given redshift interval
where their faintest observable absolute magnitudes are
$M^{Ref}_{faint}(z_{low}, {\rm SED}_1) < M^{Ref}_{faint}(z_{low}, {\rm
SED}_2)$.
\begin{figure}
\includegraphics[width=8.5cm]{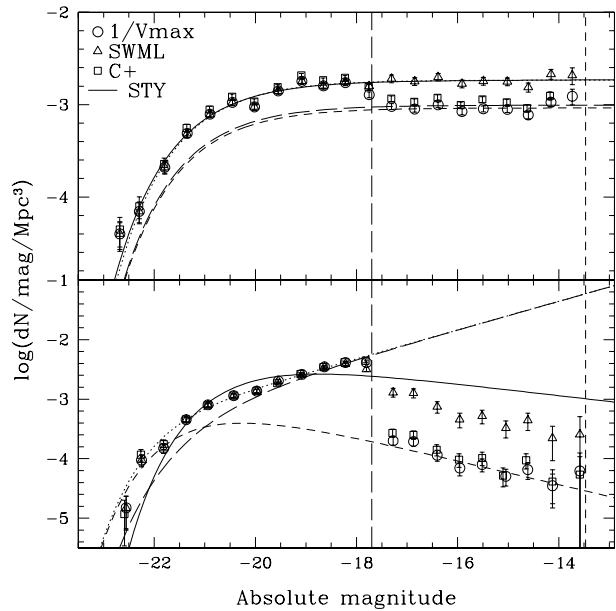} 
\caption{We simulate a simple mock catalogue with two SEDs of galaxies. 
  SED$_1$ late-type galaxy population has a faint observable absolute
  magnitude limit at $-17.7$ mag (long-dashed vertical line) and an
  input LF shown by the long-dashed curved line. SED$_2$ early-type
  galaxy population has a fainter observable limit at $-13.5$ mag
  (short-dashed vertical line) and an input LF shown by the
  short-dashed curved line.  The input global simulated LF is the
  dotted curved line, which is the sum of the input LFs of SED$_1$ and
  SED$2$.  In the top panel, the input LFs for SED$_1$ and SED$_2$ are
  the same.  The 1/V$_{\rm max}$ LF estimate (circles) and the C$^{+}$
  LF estimate (squares) do not recover the input global LF at
  magnitudes fainter than $-17.7$ mag where SED$_1$ is no more
  observable, and thus the estimate is equal to the SED$_2$ LF. The
  SWML LF estimate (triangles) and the STY estimate (solid line)
  recover the input global LF since their shapes are constrained by
  the shape of SED$_2$ input LF only, which is the same as SED$_1$
  input LF. In the bottom panel, the input LFs for SED$_1$ and SED$_2$
  have a faint-end slope $\alpha = -1.6$ and $\alpha = -0.5$
  respectively.  At magnitudes fainter than $-17.7$ mag, where SED$_1$
  population is no more detected, the 1/V$_{\rm max}$ LF and the
  C$^{+}$ LF are equal to SED$_2$ input LF. The SWML and STY LF
  faint-ends (constrained by the shape of SED$_2$ input LF) are
  underestimated.}
\label{fig3}
\end{figure}
The global LF is well recovered in the absolute magnitude range
$]-\infty, M^{Ref}_{faint}(z_{low}, {\rm SED}_1)]$ where both SEDs are
observable.  For $M^{Ref}_{faint}(z_{low}, {\rm SED}_1) < M^{Ref}_i
\leq M^{Ref}_{faint}(z_{low}, {\rm SED}_2)$, the shape of the LF is
only constrained by the probability of SED$_2$ population, since
SED$_1$ population is not detected in this range of absolute
magnitudes.  Thus the global LF has the same shape as the SED$_2$ LF.
In the range $]M^{Ref}_{faint}(z_{low}, {\rm SED}_1), +\infty[$, the
SWML and STY methods estimate the shape of the SED$_2$ LF rather than
the shape of the global LF.  Three cases may occur as follows.  First,
the LF of SED$_1$ and SED$_2$ populations have the same shape; the
global LF is well recovered.  Second, the SED$_2$ LF has a flatter
faint-end slope; the global LF is underestimated. Third, the SED$_2$
LF has a steeper faint-end slope than SED$_1$; the global LF is
overestimated. The first two cases are illustrated in Fig.~\ref{fig3}.
In the bottom panel of Fig.~\ref{fig3}, the estimate of the LF derived
with the SWML method does not follow a Schechter function.  Thus, the
STY estimate, which assumes a Schechter parametric function and which
has the same behavior as a SWML method, is not able to recover the
input.

In conclusion, the bias in the faint-end slope of the global LF
estimated with likelihood methods depends on the LF shape of each SED.
Thus to quantify the bias, we need to use simulations which deal with
LF per SED, as we do below.

\section{APPLICATIONS TO SIMULATED AND REAL DATA}

In this section, we give a description of the Hubble Deep Field (HDF)
data and simulations (Section~4.1), then we illustrate the bias in a
qualitative way using the HDF data (Section~4.2), and finally we
quantify the bias with 1,000 simulations of the HDF survey 
and the Virmos-VLT Deep Survey (VVDS) (Section~4.3).

\subsection{Brief description of the data}

We use the public version of the multi-color mock catalogues from
\citet{arnouts04} available at {\em
  www.lam.oamp.fr/arnouts/LE\_PHARE.html}.
Simulations are based on an empirical approach u\-sing observed LFs to
derive redshift and apparent magnitude distributions.  The LFs of the
ESO-Sculptor Survey \citep[ESS;][]{delapparent} are implemented up to
redshift $z < 0.6$ for early, spiral and irregular spectral types (see
Tab.~\ref{table1}).
\begin{table}  
\begin{center}
\begin{tabular}{ccrr} \hline
Spectral class & $\Phi^* \times 10^{-3} {\rm Mpc}^{-3} $ & $\alpha$ & $M^{*}_{R}$ \\ \hline
Early    & 14.77 &    0.11 & $-$20.56               \\ 
Spiral   & 13.61 & $-$0.73 & $-$20.43               \\ 
Irr      & \hspace{0.1cm}6.52 &  $-$1.64 & $-$19.84               \\ \hline
\end{tabular}
\end{center}
\caption{Input LF parameters per spectral class in the R band 
(in Vega system) used for the simulations in Section~4.} 
\label{table1}
\end{table} 
The LF evolution per type beyond $z \sim 0.6$ is constrained in a way
that it reproduces the observed redshift, number count and color
distributions.  We made the following assumptions as in
\citet{arnouts04}: the LFs of early and spiral class are constant with
redshift; the faint-end slope for the three LFs is constant; for the
irregular class there is an evolution in density in the redshift range
$0.15 < z \le 0.7$ using $\phi^*(z)=\phi^*(1+3.69(z-0.15))$, and an
evolution in luminosity in the redshift range $1.25 < z \le 2.5$ using
$M^*(z)=M^*-0.12(z-1.25)$. 
Mock catalogues are derived using the set
of SEDs as described in \citet{1999MNRAS.310..540A}, composed of 72
SEDs that have been interpolated between four observed spectra of
\citet{1980ApJS...43..393C} and two starburst models, computed with GISSEL
\citep{1993ApJ...405..538B}.  The set of SEDs is divided into three
main spectral classes: elliptical, spiral and irregular galaxies.
Since a spectral class corresponds to several different SEDs, the bias
may arise also within a single class. To test this last point, we also
use only one SED per spectral class, even though the simulation with a
set of only three SEDs is less realistic.

We used the photometric catalogue and the photometric redshifts of the
HDF South and North surveys from \citet*[1999,][]{2002MNRAS.329..355A}.  
Our aim is to illustrate the bias with observational data, and it is
not to derive the best LF of the HDF.
The HDF survey is taken as an example of a survey in which,
because of the relatively small number of observed galaxies,
only the global LF, rather than the LF for different galaxy types,
can be reliably derived in different redshift bins.
For consistency in the comparison, the HDF data have been
analysed using exactly the same set of 72 SEDs as in the simulation.

\subsection{Qualitative description of the bias}

\begin{figure*}
\begin{minipage}[htbp]{150mm}{
\includegraphics[width=150mm]{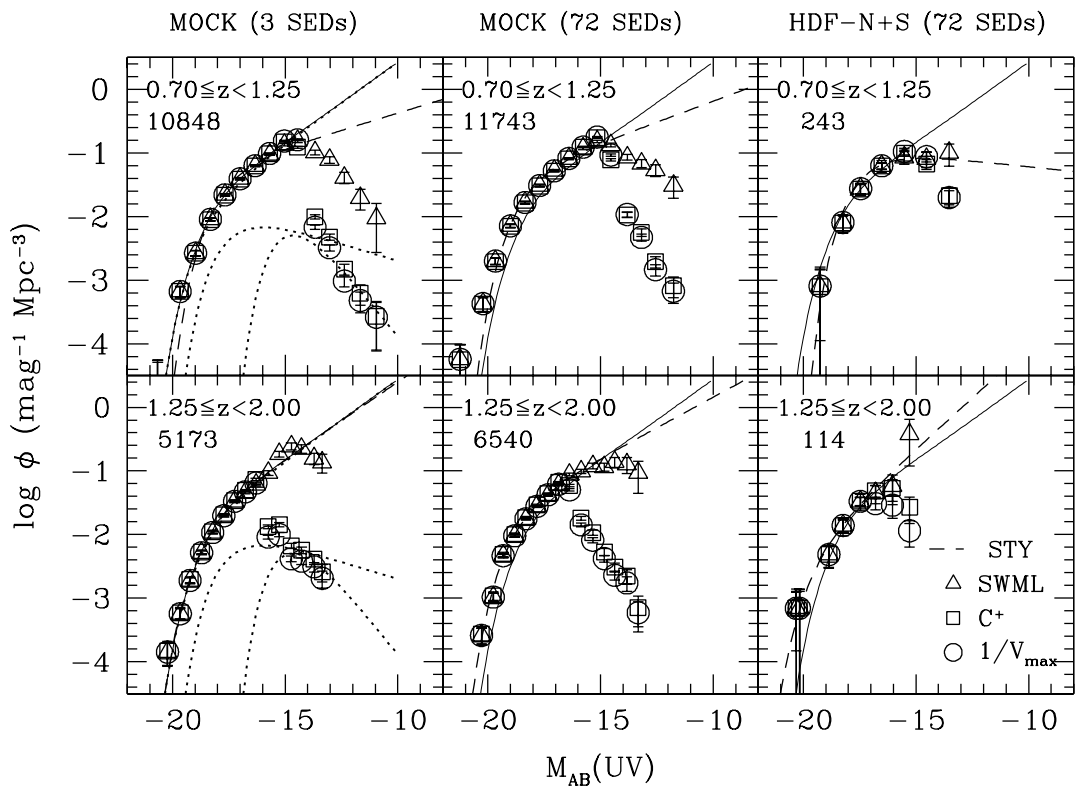} }
\end{minipage}
\caption{\hspace*{-0.08cm}a Global luminosity functions derived in the UV-FOCA
  (2000~\AA) reference filter in two redshift intervals, $0.70 \leq z < 1.25$
  (top panels), $1.25 \leq z < 2.0$ (bottom panels). The left and middle
  panels correspond to the LFs derived from simulations with 3 and 72 SEDs
  respectively.  The right panels correspond to the LFs derived from the
  HDF-North and -South surveys. The limiting magnitude in all cases is
  $I_{AB}(8140$~\AA$) < 26$ mag.  In the left panels, we display the LFs
  corresponding to the three input SEDs used (dotted lines): from the steepest
  to the shallowest slope, it is the irregular-, spiral- and elliptical-type
  LF respectively.  In each panel, we plot the global simulated LF (solid
  line) corresponding to the sum of the three input LFs. We plot also
  the results from the following global LF estimates, STY (dashed line), SWML
  (triangles), C$^{+}$ (squares), and 1/V$_{\rm max}$ (circles). Below the $z$
  redshift intervals we quote the number of galaxies used to derive the LFs.
  We adopt Poissonian error bars for the 1/V$_{\rm max}$ and C$^{+}$
  estimators. The error bars for the SWML estimator are derived
  following the EEP88 method.
}
\label{fig4a}
\end{figure*}

We describe the bias using HDF data and one simulation from which data
have been selected with the same apparent magnitude limits as the HDF
data. To illlustrate the bias in a qualitative way, we use a large
number of galaxies ($> 5,000$ in each redshift bin) so that
statistical uncertainties become ne\-gli\-gi\-ble.  Moreover we did
not include any surface brightness effect and any uncertainties on
apparent magnitudes, redshifts, and absolute magnitudes.  We derived
the absolute magnitudes and the global LFs in two redshift intervals:
$0.70 \le z < 1.25$ and $1.25 \le z < 2.00$ in the following reference
filters, UV (Fig.~\ref{fig4a}a), B (Fig.~\ref{fig4b}b) and I
(Fig.~\ref{fig4c}c) with samples selected at $I_{AB} \le 26$ mag.  In
these figures, we plot the input LFs of the simulation (one for each
spectral type and the global one) and the estimates of the global LF
derived by the 1/V$_{\rm max}$, C$^{+}$, SWML and STY methods from
mock catalogues and the HDF data.  Before discussing in detail each
figure, we note that in all cases the estimators give similar results
using the mock catalogues constructed with three SEDs only and those
with 72 SEDs.
\setcounter{figure}{3}
\begin{figure*}
\begin{minipage}[htbp]{150mm}{
\includegraphics[width=150mm]{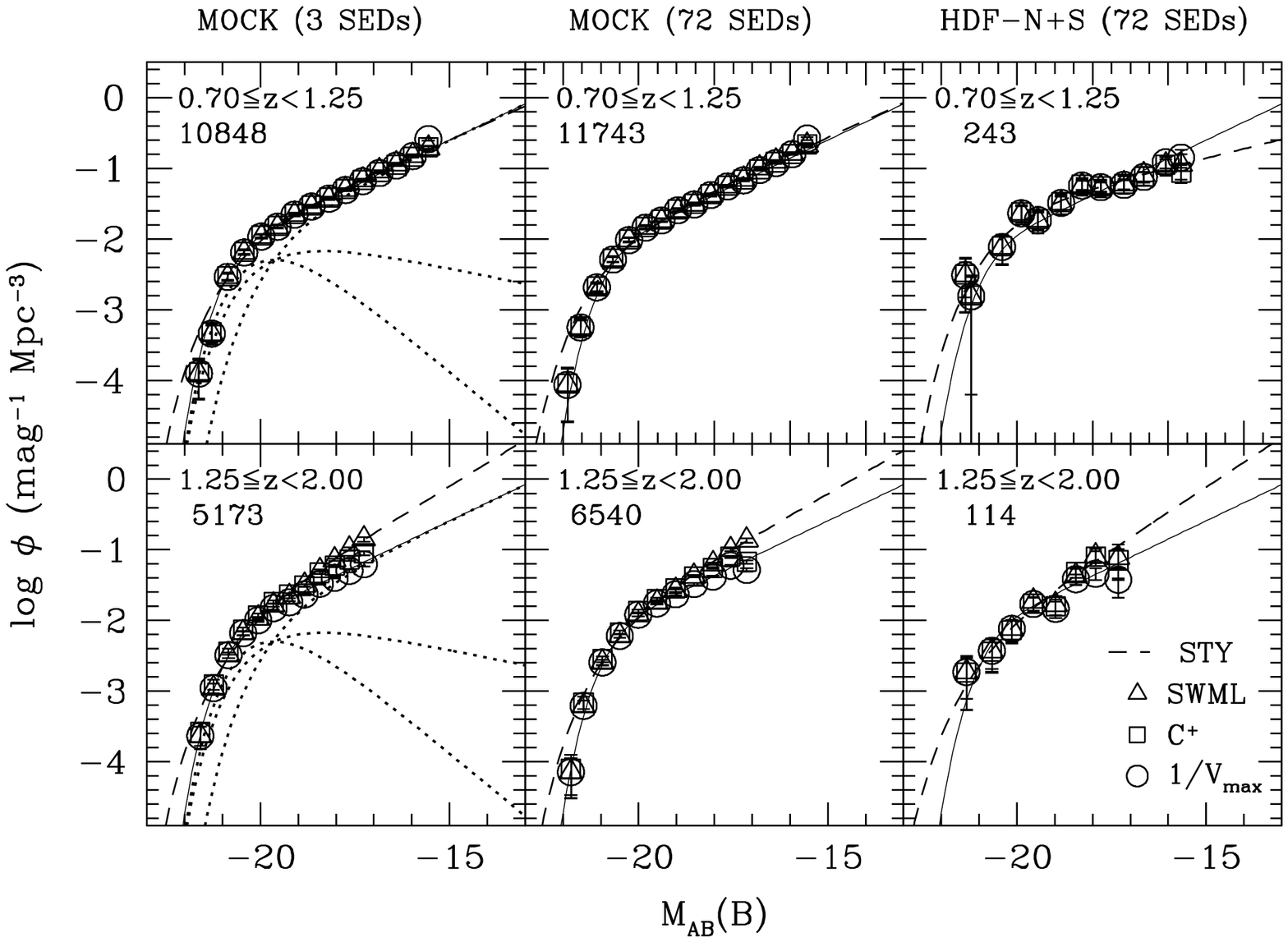} }
\end{minipage}
\caption{\hspace*{-0.08cm}b Same as Fig.~\ref{fig4a}a except that the reference filter is B-HST (4500~\AA).}
\label{fig4b}
\end{figure*}
\setcounter{figure}{3}
\begin{figure*} 
\begin{minipage}[htbp]{150mm}{
\includegraphics[width=150mm]{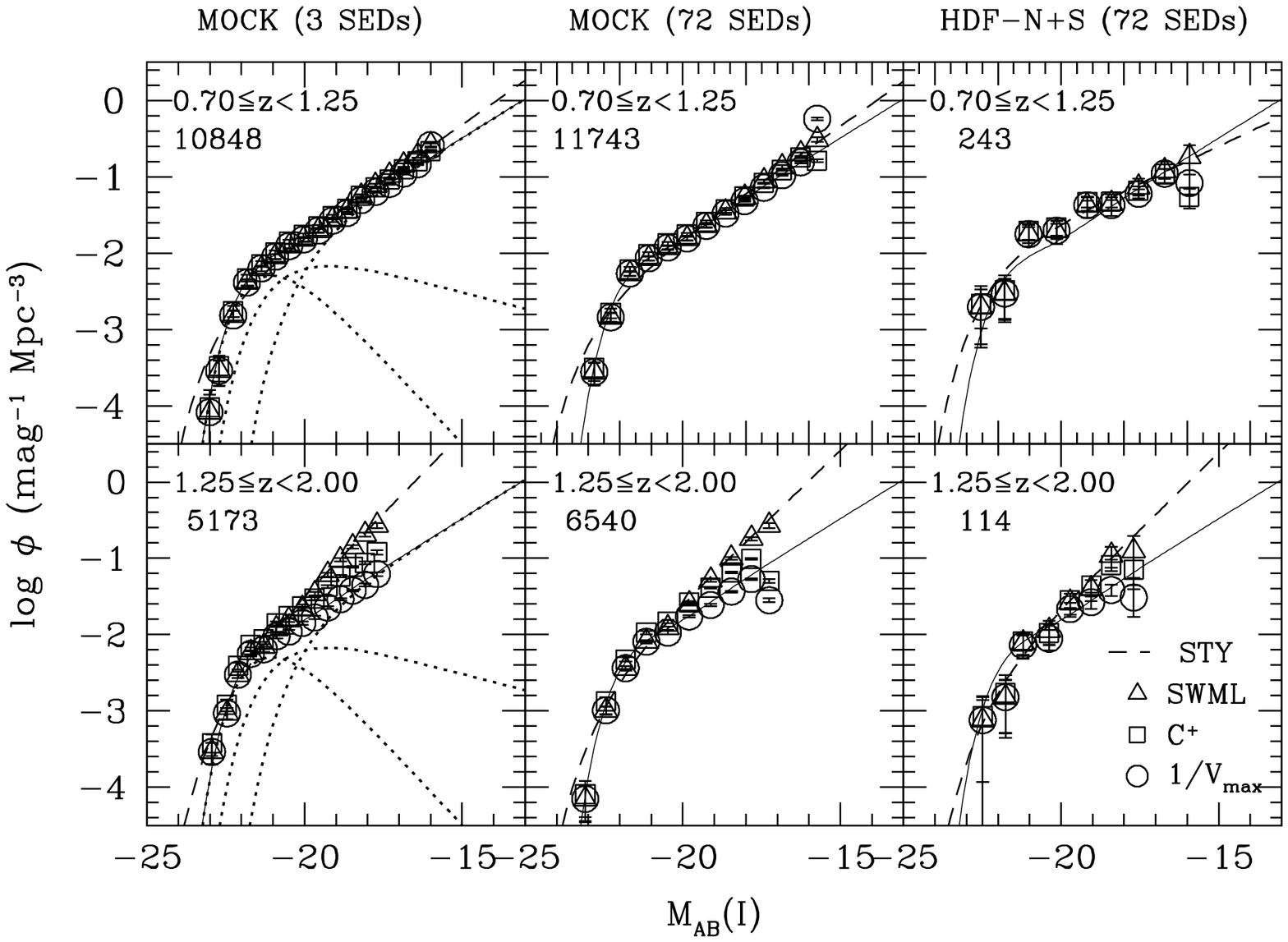} }
\end{minipage}
\caption{\hspace*{-0.08cm}c Same as Fig.~\ref{fig4a}a except that the reference filter is I-HST (8140~\AA).}
\label{fig4c}
\end{figure*}
We display the simulations with three SEDs so to compare exactly to
the three input simulated LFs per class (i.e. per SED in this case).
With 72 SEDs such a comparison is less obvious since each class
contains several SEDs.  However we show also the simulations with 72
SEDs so it is comparable to the analysis applied to the HDF data, that
is with the same selection
function, the same set of SEDs and the same filters. \\

\subsubsection{$UV$-rest LFs}

In the left panels of Fig.~\ref{fig4a}a, we show the individual input LFs
(dotted lines) of the mock catalogue built with three SEDs, that is with one
SED per spectral class. The sum of the three individual input LFs corresponds
to the global simulated LF (solid line). Thus the LF estimators should be able
to recover this global LF in estimating the LF for the whole galaxy sample.
First, we note that all estimators underestimate the faint-end of the LFs. As
previously shown in Fig.~\ref{fig2}, the irregular-type galaxies `disappear'
from the faintest absolute ma\-gni\-tude bins (for $M^{UV} > -14.2$ mag at
$z_{low} = 0.7$ and $M^{UV} > -16.0$ mag at $z_{low} = 1.25$).  As explained
in Section~3 when one type `disappears' the global LF estimate is based only on
the remaining types. 

Left panels of Fig.~\ref{fig4a}a show this point; 
the 1/V$_{\rm max}$ and
C$^{+}$ estimators recover the input LF of elliptical galaxies in the faintest
absolute magnitude bins (i.e. open circles and squares overlap the dotted line
of the elliptical-type LF).  The SWML method underestimates the faint-end
slope of the global LF since this estimator recovers the shape of the
elliptical-type galaxies (i.e.  open triangles follow the same slope as the
elliptical-type LF).  The less biased result is obtained with the STY method
and the more biased results are obtained with the 1/V$_{\rm max}$ and C$^{+}$
methods.  It is interesting to note in the redshift interval $1.25 \leq z <
2.0$ that the little `bump' of the global SWML LF estimate at $M^{UV} \sim
-15$ mag corresponds to the shape the elliptical-type input LF.

In the middle panels, we show the global LF derived using the mock
catalogue with 72 SEDs distributed into three spectral classes.  We see that
the increase of the number of SEDs used per spectral class does not affect
significantly the trend of the LF estimators described above. Using many
SEDs smooths the effect of the `disappearance' of a SED.

In the right panels, we show the global LF derived from the HDF data
using the same 72 SEDs as in the mock catalogue.  We do not observe
very faint (and very bright) galaxies as observed in the simulation
even though both data sets have the same selection function. We have
tested that this is indeed due to the relatively small number of
galaxies observed in the HDF and the small probability to observe the
faintest (and brightest) galaxies.  Thus, the difference between the
various estimators is less obvious than in the simulation. However we
note here the same general trend in the faintest observed bins.  This
may suggest that the bias which affects the global HDF LF estimation
is similar to the bias shown in the simulation. In conclusion the
faint-end of the global LF cannot be properly recovered by any
estimator as can be seen from the comparison with the global simulated
LF (solid line).

\subsubsection{$B$-rest LFs}

We plot in Fig.\ref{fig4b}b the global LF in the $B$ reference filter with the
same symbols as in Fig.\ref{fig4a}a. The simulations built with both three and
72 SEDs give exactly the same result. At $z = 0.7$, the $B$-filter absolute
magnitudes are almost independent on SEDs since the selection filter ($I$)
roughly corresponds to the $B$-rest filter at this redshift. As a consequence,
all estimators recover very well the global simulated LF in the first redshift
bin ($0.70 \leq z < 1.25$).  For the HDF data, the four estimators are in
agreement amongst them.  The bias is not present in this case, as shown by our
simulation.

In the redshift bin $1.25 \leq z < 2$, the SWML and the STY methods
slightly overestimate the slope of the global LF.  At $z_{low} =
1.25$, the limiting absolute magnitude $M_{faint}^B$ for early types
is about one magnitude brighter than for late types.  Thus the SWML
and STY estimators measure the steep slope of the late-type galaxies,
and this explains the overestimate of the faint-end slope of the
global measured LF. The 1/V$_{\rm max}$ and C$^{+}$ methods well
recover the global simulated LF, since the density of the first type
which `disappears' from the sample (that is the faint elliptical-type
galaxies) is negligible in the global simulated LF. Global LF
estimators from the HDF data are globally in agreement amongst them.
The estimate of the global LF in the $B$ reference filter is quite
robust.

\subsubsection{$I$-rest LFs}

For the $I$-rest LFs, the range of absolute magnitudes in which different
spectral types `disappear', increases with redshift since the reference filter
is also the filter used to select the sample.  At $z = 0.7$, this range spans
one ma\-gni\-tude (see Fig.~\ref{fig2}), thus all estimators should be slightly
biased in the redshift bin $0.7 \leq z < 1.25$. We note that all LF estimators
recover the global simulated LF in simulations derived with three as well as
with 72 SEDs. For the HDF LF, the estimators are in good agreement with each
other and this suggests that the measurement is not biased.

At $z = 1.25$, the absolute magnitude range in which different
spectral types `disappear' spans more than two magnitudes. In this
case the first spectral types which `disappear' from the faintest bins
are the elliptical types.  In both simulations, the SWML and STY
methods largely overestimate the slope of the faint-end in the
redshift bin $1.25 \leq z < 2.00$, while the 1/V$_{\rm max}$ and
C$^{+}$ methods recover reasonably well (the 1/V$_{\rm max}$ method
better than the C$^{+}$ method) the faint-end of the global LF.
Indeed, the slope estimated with the SWML and STY methods is the slope
of the irregular galaxies, which is steeper than the global input LF
slope in the range of absolute magnitudes here measured.  The
1/V$_{\rm max}$ and C$^{+}$ estimators recover the slope of the global
LF since the contribution of the faint elliptical-type galaxies to the
global LF is negligible.  In the global LF derived from the HDF data,
the SWML and STY estimators predict a steeper slope than the 1/V$_{\rm
  max}$ and C$^{+}$ estimators, as in the simulations. According to
the simulations, the global LF derived from the HDF data and estimated
with the SWML and STY overestimate the faint-end slope in
the redshift bin $1.25 \leq z < 2.00$.

\subsection{Quantitative measurement of the bias}

As shown throughout the paper, the bias depends on the band in which
galaxies are selected, on the apparent limits of the survey, on the
reference band in which the LF is measured, and on the lower limit of
a studied redshift bin.  Thus we cannot provide here a recipe to
quantify the amplitude of the bias in all cases. We choose to quantify
it in the following specific cases, the HDF and the Virmos-VLT Deep
Survey (VVDS) surveys.  The HDF is a good example where we can derive
the global LF only.  The VVDS is a deep spectroscopic survey with
enough galaxies to derive the LFs per type \citep{LeFevre04}.  The
global LF is also a necessary output of the VVDS survey, however even
in this case, deriving it from the sum of extrapolated LFs per type
may be dangerous.  We produced 1,000 simulations representative of HDF
and VVDS surveys.  Realistic apparent magnitude errors are introduced
in each simulation, and this has resultant uncertainties on the
template fit, k-corrections and absolute magnitudes. We do not include
any surface brighness effects and errors on the redshift.

We choose the Schechter parameters to quantify the bias.  The
parameter $\alpha$ is the most affected by the bias.  In
consequence, we fix $M^*$ for the LF estimation of each simulations.
We call $\Delta \alpha$, the difference between the estimated $\alpha$
value and the input $\alpha$ value to the simulations and 
$\overline{\Delta \alpha}$ the average value of $\Delta \alpha$ over
1,000 realizations.

In the simulations, we have implemented LFs per type.  Then to compare
the estimates of the global LF with the input, we had to define a
'pseudo' input global LF. We did that in the following way.
\begin{itemize}
\item We realize one big simulation with more than 1,000
  galaxies in each redshift bin so to minimize statistical
  fluctuations.
\item We do not include any errors on redshifts, apparent
   magnitudes, absolute magnitudes, and types.
\item To obtain a global LF which is not affected by the bias 
  in the big simulation, we select galaxy samples 
  three magnitudes fainter than the apparent magnitude limits 
  of the 1,000 simulations representative of HDF and VVDS. 
  And then the absolute magnitude range is limited to
  the same absolute magnitude range spanned by the LF estimates 
  of these 1,000 realizations. The Schechter parameters are 
  derived with the STY method. 
\end{itemize}

\subsubsection{Bias quantified for HDF}

\begin{figure*}
\begin{minipage}[htbp]{150mm}
{\includegraphics[width=150mm]{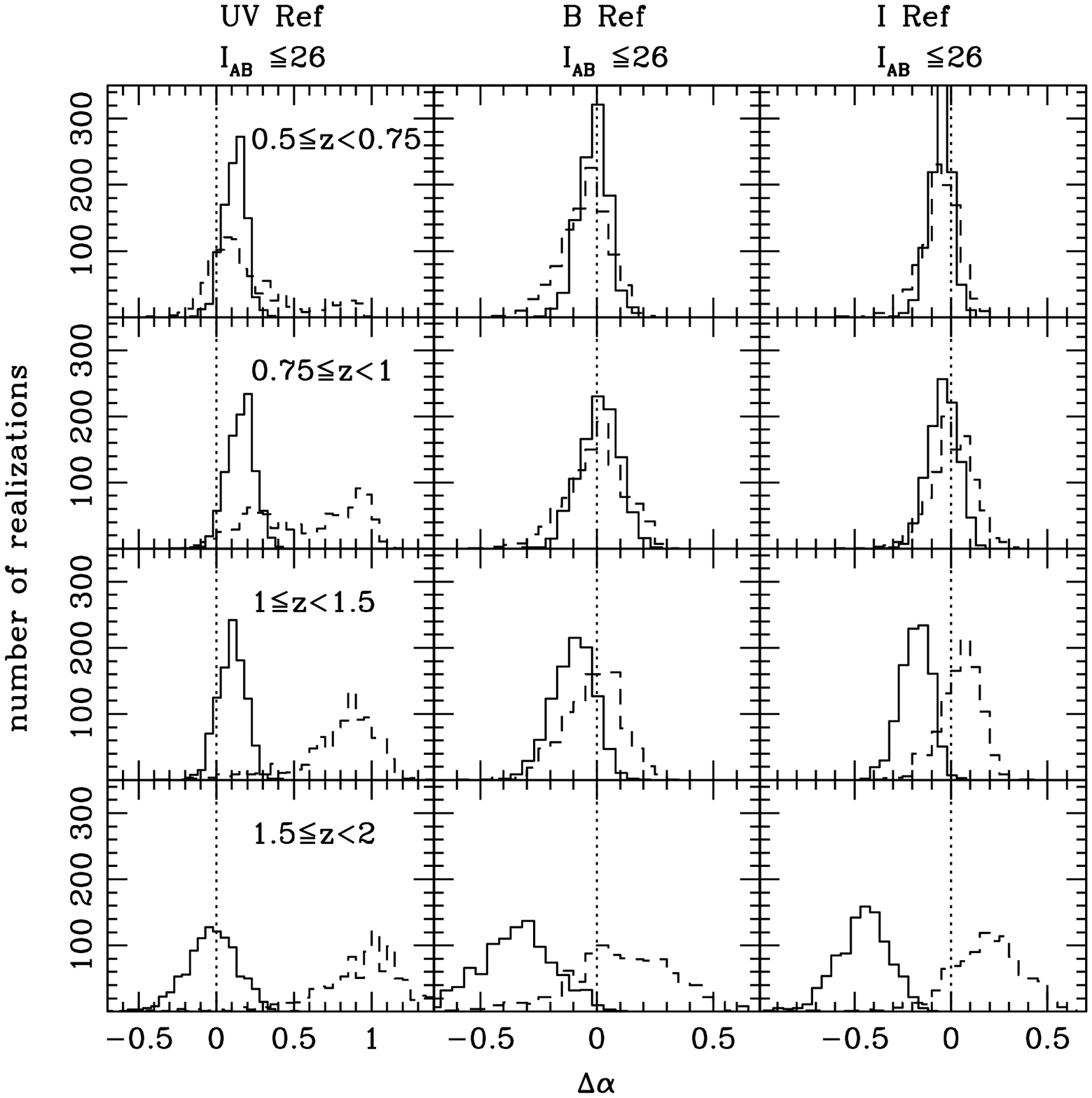} }
\end{minipage}
\caption{\hspace*{-0.08cm}{Histograms of the differences, $\Delta\alpha$, 
    between the estimated and the input $\alpha$ parameters, over
    1,000 realizations done with the HDF characteristics.
    $\Delta\alpha$ for the STY estimates is the solid line histogram,
    while $\Delta\alpha$ for the 1/V$_{\rm max}$ fit is the
    dashed-line histogram.  All the simulated HDF samples are selected
    with $I_{AB}\le 26$.  Panels in the first column correspond to the
    case with the global LFs derived in the UV-FOCA (2000~\AA) filter,
    in the middle column in the B-HST (4500~\AA) filter, and in the
    right column in the I-HST (8140~\AA) filter.  From top to bottom
    panels, $\Delta\alpha$ is measured within the redshift bins [0.5, 0.75], 
    [0.75, 1], [1, 1.5], [1.5, 2].}}
\label{fig5}
\end{figure*}

Each simulation of the HDF is realized on 8 arcmin$^2$ and with
$I_{AB} \le 26$.  The LFs are derived in three reference bands
(UV-FOCA, B-HST and I-HST) and in four redshift bins from $z=0.5$ to
$z=2$. The results of these simulations are shown in Fig.~\ref{fig5}.

In the UV~(2000~\AA) band (first column of the figure), the estimate
of the faint-end slope with the 1/V$_{\rm max}$ fit is
strongly underestimated ($\overline{\Delta \alpha} > 0.5$ for $z \ge
0.75$), while the STY estimate is only slightly biased
($\overline{\Delta \alpha} \simeq 0.1$).  In the B-HST band (second
column), the LF estimates derived with the STY and the 1/V$_{\rm max}$
methods are robust up to the redshift bin [1, 1.5].  In the redshift
bin [1.5, 2], the STY estimate gives $\sim -0.30$ steeper values for the
$\alpha$ parameter, whereas the 1/V$_{\rm max}$ fit gives $\sim 0.20$
flatter values.  In the I band (third column), the bias from the STY
estimate increases as a function of redshift; the input $\alpha$
parameter is recovered in the first redshift bin, and
$\overline{\Delta \alpha}$ decreases down to $-0.40$ in the last
redshift bin [1.5, 2].  The 1/V$_{\rm max}$ fit gives an estimate
which recovers the input $\alpha$ up to the reshift bin [1, 1.5].

\subsubsection{Bias quantified for VVDS}

We quantify now the amplitude of the bias for a sample selected in a
way similar to the VVDS. We adopt the Johnson Kron-Cousins filter set
and we perform each simulation on 360 arcmin$^2$. The sample selected
from each simulation is approximatively of the same size ($\sim 7,000$
galaxies) as the presently available spectroscopic deep sample in the
0226-0430 VVDS field.  The global LF estimations are derived in three
reference bands, U (3600~\AA), B (4200~\AA) and I (8000~\AA), and in
four redshift bins from $z=0.6$ to $z=1.5$. 
For our illustration, we consider two cases of sample selection; an 
I-selected sample (as the VVDS spectroscopic data), and 
an U-selected sample (as for instance, a VVDS photometric data).  

\begin{figure*}
\begin{minipage}[htbp]{150mm}
{\includegraphics[width=150mm]{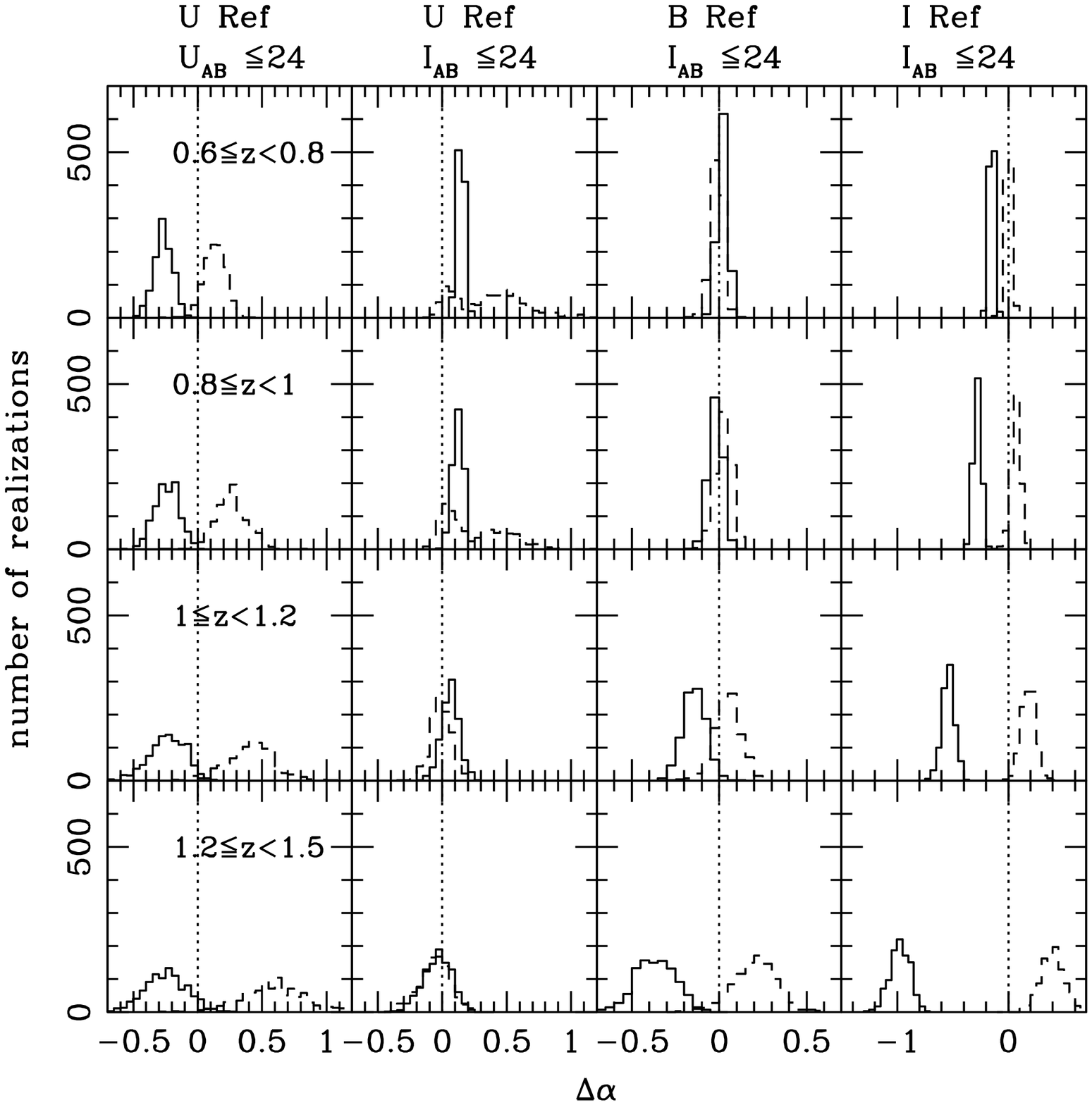} }
\end{minipage}
\caption{\hspace*{-0.08cm}{
    Same as in Fig.~\ref{fig5} except that the simulations are
    selected in a way similar to the VVDS.  Panels in the first column
    correspond to samples selected with U$_{AB}$ $\le 24$, and the
    other columns correspond to samples selected with $I_{AB} \le 24$.
    In the first and second column, the global LFs are derived in the
    U~(3600~\AA) reference filter, the third column in the
    B~(4300~\AA), and the last column in the I~(8000~\AA). From top to
    bottom panels, $\Delta\alpha$ is measured within the redshift bins
    [0.6, 0.8], [0.8, 1], [1, 1.2], [1.2, 1.5].}}
\label{fig6}
\end{figure*}

The results of these simulations are shown in the Fig.~\ref{fig6}. We
show in the first column the U-band LF estimates for samples selected
in the U band with $U_{AB}\le 24$. 
The bias increases
with redshift. In all redshift bins, the STY estimate gives steeper
faint-end slopes ($\overline{\Delta \alpha} \sim -0.3$), and the
1/V$_{\rm max}$ fit gives flatter faint-end slopes ($\overline{\Delta
  \alpha} > 1$ for $z \ge 1$).

In the second column, samples are selected with $I_{AB}\le 24$, and
the LF estimates are derived in the U band.  Like for the HDF, the
faint-end slope is underestimated by both the STY method and
the 1/V$_{\rm max}$ fit in the first redshift bin
($\overline{\Delta \alpha} \sim 0.2$ and $\overline{\Delta \alpha}
\sim 0.5$ respectively). The bias decreases with redshift.  In the
last redshift bin, all estimates recover the input $\alpha$ value
because the rest-frame selection I-band corresponds to the reference
U-band at $z \sim 1.2$.

In the third column, samples are selected with $I_{AB}\le 24$, and the
LF estimates are derived in the reference B-band.  The input $\alpha$
parameter is recovered by both methods up to $z=1$.  The STY estimation of
$\alpha$ in the redshift bin [1, 1.2] is only slightly biased
($\overline{\Delta \alpha}\sim -0.1$), while it is strongly biased in
the last redshift bin [1.2, 1.5].  The STY estimate gives steeper
faint-end slopes ($\overline{\Delta \alpha}\sim -0.4$), and the
1/V$_{\rm max}$ fit gives flatter faint-end slopes
($\overline{\Delta \alpha}\sim0.2$).

In the last column, samples are selected with $I_{AB}\le 24$, and the
LF estimates are derived in the reference I-band.  The bias
increases with redshift. The 1/V$_{\rm max}$ fit gives flatter
faint-end slopes (up to $\overline{\delta \alpha}\sim0.4$ for $1.2\le
z < 1.5$).  The STY estimate gives steeper faint-end slopes
(up to
$\overline{\Delta \alpha}\sim -1$ for $1.2\le z < 1.5$).\\

In conclusion, in deep surveys like the HDF or the VVDS, the various
estimators fail to correctly recover the global LF faint end if the
reference filter is far from the rest-frame selection filter.  If the
results from different estimators are not in good agreement with each
other, the global LF estimate is likely to be biased and
the only way to quantitatively estimate this bias is through a
detailed comparison with simulations representing to the sample
properties.

\section{CONCLUSIONS}

Our study enabled us to describe when LF estimators are robust for the
measurement of the global LF in the framework of the earlier and
future deepest surveys. We demonstrated that the estimation
of the global LF contains an intrinsic bias due to the fact
that, in a magnitude limited sample, different galaxy types have
different limits in absolute magnitude because of different
k-corrections. The importance of the effect is larger when the range
of k-correction between the different galaxy types is wide. For this
reason this bias mainly arises in high redshift samples.  The STY and
SWML estimators are not affected in the same way by this bias as the
1/V$_{\rm max}$ and C$^{+}$ estimators.  If the STY, SWML and the
1/V$_{\rm max}$, C$^{+}$ methods are not in good agreement with each
other, this is an indication that the bias in the global LF estimators
is present. A good indication of the presence of a significant bias is
when the differences between different estimators (Vmax and STY for
instance) is larger than the statistical uncertainties (Poisson errors
for instance).  We quantified it using realistic simulations and
observations for galaxies selected in the $I$ filter, and measuring
the LF in various reference filters (UV, B, I). We obtain the
following results.
\begin{itemize}
\item[(i)] Case $1+z_{low} < \lambda^S/\lambda^{Ref}$ (e.g., a reference-frame
  UV LF for galaxies selected in I): the studied estimators underestimate the
  faint-end slope of the global LF for $z_{low} \la 2 $. This underestimate is
  particulary significant for the 1/V$_{\rm max}$ and C$^{+}$ methods
  (i.e. for instance, the $UV$-LF of the SDF).
\item[(ii)] Case $1+z_{low} \sim \lambda^S/\lambda^{Ref}$ (e.g.,
  a reference-frame B LF for galaxies selected in I): the
  estimators of the global LF are robust up to $z_{low} \la 1.3$. In this
redshift range the bias is minimal (i.e. for instance, the CFRS
  case).
\item[(iii)] Case $1+z_{low} > \lambda^S/\lambda^{Ref}$ (e.g.,
  a reference-frame I LFs with galaxies selected in I): the STY
  and SWML methods overestimate the faint-end slope of the global LF, while
  the 1/V$_{\rm max}$ method roughly recovers well the global LF
  (e.g., for instance, the redshift bin [1.25, 2] of the HDF).
\end{itemize}

The ways to reduce the intrinsic bias of the global LF
estimators are the following:
\begin{itemize} 
\item[(a)] The selection of galaxy subsamples in the closest
  rest-frame filter to the reference filter in which the LF is
  measured \citep[see e.g.][2003]{2001ApJ...551L..45P}.  This method is
  also the best to reduce the SED dependency in the measurement of
  absolute magnitudes since in this case the term
  [color+$k$-correction] is little dependent on the SED.  Only
  multi-color surveys allow to derive the same rest-frame band LF at
  different redshifts using this strategy.
\item[(b)] In principle, the estimate of the global LF using the sum of the
  extrapolated LF per galaxy type.  However it requires a good
  knowledge of the slope for all the LFs per type, and in practice 
  the use of extrapolated LFs may be hazardous.
\item[(c)] The estimate of the global LF using a filter in which the
  differences between $k$-corrections are small, as for instance in
  the $K$-filter, e.g. \cite{2002A&A...395..443B}, \cite{poz}.
\item[(d)] The estimate of the global LF within an absolute magnitude
  range in which all galaxy types are detected \citep[see
  e.g.][]{1997ApJ...487..512S}. This method is appropriate for very
  large surveys like the VVDS for instance, at the cost of the loss of
  the faintest bins of the global LF.
\end{itemize}

\section*{Acknowledgments}

This work has been developed within the framework of the VVDS Consortium. 
We thank the referee for the careful reading of the manuscript and 
the useful suggestions.

\label{lastpage}
\end{document}